\documentclass{pasj00}
\usepackage{color}
\draft

\begin{document}
\SetRunningHead{Y. Takeda and S. Honda}{Oxygen abundances of M~67 stars 
}
\Received{2014/10/08}
\Accepted{2014/12/18}

\title{On the oxygen abundances of M~67 stars\\
from the turn-off point through the red-giant branch
\thanks{Based on data collected by using the NAYUTA Telescope
of the Nishi-Harima Astronomical Observatory.}
}

%

\author{
Yoichi \textsc{Takeda}\altaffilmark{1,2} and
Satoshi \textsc{Honda}\altaffilmark{3}
}

\altaffiltext{1}{National Astronomical Observatory, 2-21-1 Osawa, 
Mitaka, Tokyo 181-8588}
\email{takeda.yoichi@nao.ac.jp}
\altaffiltext{2}{The Graduate University for Advanced Studies, 
2-21-1 Osawa, Mitaka, Tokyo 181-8588}
\altaffiltext{3}{Nishi-Harima Astronomical Observatory, Center for Astronomy,\\
University of Hyogo, 407-2 Nishigaichi, Sayo-cho, Sayo, Hyogo 679-5313
}
%


\maketitle

\begin{abstract}
With an aim to examine whether the surface oxygen composition suffers any appreciable 
change due to evolution-induced mixing of nuclear-processed material in the envelope 
of red giants, abundance determinations for O/Fe/Ni based on the synthetic 
spectrum-fitting method were performed by using the moderate-dispersion spectra 
in the 7770--7792$\rm\AA$ region (comprising O~{\sc i} 7771--5, Fe~{\sc i} 7780, 
and Ni~{\sc i} 7788 lines) for 16 stars of the old open cluster M~67 in various 
evolutionary stages from the turn-off point through the red giant branch. 
We could not find any meaningful difference in the oxygen abundances between 
the non-giant group ($T_{\rm eff} > 5000$~K) and the red-giant group ($T_{\rm eff} < 5000$~K),
which are almost consistent with each other on the average (despite that both have 
rather large dispersions of a few tenths dex caused by insufficient data quality),
though only one giant star (S~1054) appears to show an exceptionally low O abundance 
and thus needs a more detailed study. 
This result may suggest that oxygen content in the stellar envelope is hardly 
affected (or changes are insignificant) by the mixing of H-burning products 
in the red-giant phase, as far as M~67 stars of low mass ($\sim 1.3 M_{\odot}$) 
are concerned, which is consistent with the prediction from the conventional 
stellar evolution theory of first dredge-up.
\end{abstract}

%


\section{Introduction}

As a star is evolved off the main sequence after exhaustion of
hydrogen fuels in the core, it increases its radius while the 
surface temperature drops down, and the deep convection zone is 
developed. As a result, some portion of the H-burning (CNO cycle) 
product in the interior may be salvaged and mixed to alter 
the surface abundances of red giants. According to the canonical 
stellar evolution calculation, it is essentially the CN-cycled 
(C$\rightarrow$N reaction) material that is dredged up, 
while the product of ON-cycle (O$\rightarrow$N reaction; occurring 
in deeper region of higher $T$) is unlikely to cause any significant
abundance change because mixing is not expected to substantially 
penetrate into such a deep layer; thus the predicted surface abundances 
are characterized by a deficit in C as well as an enhancement in N
(with its peculiarity degree increasing with mass/luminosity), 
while O is practically unaffected (see, e.g., Fig. 24 in Mishenina et al. 2006).

While such expected tendency has been almost confirmed observationally 
for C and N, the behavior of O is still controversial and unsettled:\\
--- Mishenina et al. (2006) reported that the oxygen abundances 
(determined based on the [O~{\sc i}] forbidden line at 6300~$\rm\AA$)
in red-clump giants of $\sim$~1--3~$M_{\odot}$ are almost normal 
($\langle$[O/Fe]$\rangle \simeq 0$), in agreement with the theoretical 
prediction.\\
--- Further, Tautvai\u{s}ien\.{e} et al.'s (2010) similar study
on red-clump giants resulted in essentially the same conclusion as
Mishenina et al.'s.\\
--- However, Takeda, Sato, and Murata (2008) reported in their extensive 
spectroscopic analysis of 322 late-G and early-K giants
(4500~K $\ltsim T_{\rm eff} \ltsim$ 5500~K, 1.5 $\ltsim \log g \ltsim$ 3.5,
1 $\ltsim M/M_{\odot} \ltsim$ 5) that [O/Fe] (determined from [O~{\sc i}] 5577 line) 
shows a subsolar tendency with the extent of peculiarity increasing with $M$ 
(cf. figure 12 therein), which implies that a mass-dependent 
dredge-up of (not only CN-cycle but also) ON-cycle products 
may take place in the envelope of red giants. So, if this is real,
it would require a revision of the standard theory for the dredge-up 
in the envelope of red giants. 

The best approach to check whether or not stellar surface abundances
suffer changes during the evolutionary paths from the main sequence
to the red giant phase is to investigate a number of stars (of various types) 
belonging to an open cluster. That is, since all such member stars are 
considered to have the same age as well as the same initial composition,
detection of any systematic abundance differences along the evolutionary 
sequence would make a direct evidence for the existence of evolution-induced
build-up of chemical peculiarity.
Above all, the well known old open cluster M~67 (NGC~2682) is especially 
suitable for this purpose, since (unlike many other comparatively young 
clusters) it has a particular merit of including a wealth of stars 
(not only on the main sequence but also) on the continuous evolutionary 
sequence from the turn-off point through the red-giant branch.

This well known old Galactic cluster has been repeatedly studied
and several reports on the oxygen abundances of M~67 member 
stars are already available; e.g., Brown (1985) for 3 giants; 
Shetrone and Sandquist (2000) for 10 blue-straggler and turn-off stars; 
Tautvai\u{s}ien\.{e} et al. (2000) for 10 giants; Yong, Carney and Teixera de Almeida 
(2005) for 3 giants; Randich et al. (2006) for 10 dwarfs/subgiants of late F- and 
early G-type; Pace, Pasquini, and Fran\c{c}ois (2008) for 5 solar-type stars;
Pancino et al. (2010) for 3 giants.\footnote{
We do not pay attention here to the pioneering studies done in 1970's 
based on photographic plates because of their less reliability; see, e.g., 
Sect. 1 of Tautvai\u{s}ien\.{e} et al. (2000).}  
However, any trial of systematic oxygen abundance studies in a
consistent manner for a wide range of M~67 stars from near-main sequence 
(F type) through red giants (K type) has never been made so far 
to our knowledge.

Motivated by this situation, we decided to conduct a spectroscopic study on  
M~67 stars in various stages from less evolved (near to turn-off 
points) through fully evolved (red giants), in order to determine 
their oxygen abundances based on O~{\sc i}~7771--5 triplet lines 
(by making use of the fact that these O~{\sc i} lines are visible 
in a wide range of stars from spectral type A through K) and to see
if any systematic oxygen abundance peculiarity (e.g., a comparative 
deficit of O in red giants) is observed. This was our primary purpose.

Somewhat disappointingly, we realized in the course of the analysis that 
the spectral data we obtained for the relevant M~67 stars of $V \ltsim 13$~mag 
with the 2-m NAYUTA telescope (used for this study) were of insufficient 
quality (i.e., S/N ratios were from only several tens to $\sim 100$) to carry 
out reliable abundance determinations. Yet, we positively considered that this is 
a good opportunity for us to learn how much abundance information is gained 
by carefully studying rather noisy spectra of medium resolution, since such 
circumstances may be encountered in astronomical spectroscopy of faint objects. 
Accordingly, we paid special attention to quantitatively estimating errors 
involved in the resulting abundances. This challenge makes another aim of 
this investigation. 

The remainder of this article is organized as follows. 
After describing our observations in section 2, 
we explain the assignments of stellar parameters in section 3. 
The procedure of our abundance determination based on the spectrum-fitting 
method is illustrated in section 4, followed by section 5 where abundance
errors are evaluated in various respects. The resulting abundances are discussed 
in section 6, and the conclusion is summarized in section 7.

\section{Observational data}

The list of our 16 targets of M~67 cluster is presented in table 1, 
which were selected to cover the evolutionary status from the turn-off point 
through the red-giant branch (cf. figure 2a).
Spectroscopic observations of these stars were carried out in 2014 January
and February using the Medium And Low-dispersion Longslit Spectrograph 
(MALLS; cf. Ozaki \& Tokimasa 2005) installed on the Nasmyth platform of 
the 2-m NAYUTA telescope at Nishi-Harima Astronomical Observatory (NHAO). 
Equipped with a 2K$\times$2 K CCD detector (13.5 $\mu$m pixel), MALLS
can record a spectrum covering $\sim$~400~$\rm\AA$ (7600--8000~$\rm\AA$)
in the medium-resolution mode with the resolving power of $R \sim 12000$. 
Since we had to limit the maximum exposure time of one frame up to 20 minutes
(because of enhanced dark level), a number of spectral frames were co-added 
in order to reduce the spectrum noise as much as possible.
The reduction of the spectra (bias subtraction, flat-fielding, 
spectrum extraction, wavelength calibration, co-adding of frames to improve S/N, 
continuum normalization) was performed by using the ``noao.onedspec'' package of 
the software IRAF\footnote{IRAF is distributed
    by the National Optical Astronomy Observatories,
    which is operated by the Association of Universities for Research
    in Astronomy, Inc. under cooperative agreement with
    the National Science Foundation.} 
in a standard manner. 

The S/N ratios of the resulting spectra were estimated in two ways.
First, assuming the photon-noise-limited case, we derived 
S/N(predicted)~$\simeq \sqrt{n_{\rm ph}}$ ($n_{\rm ph}$ is the photon counts). 
Second,  we directly measured the standard deviation ($\sigma_{k}$) and 
the average ($\langle c_{k} \rangle$) of the spectrum at each of the selected 
eight line-free windows ($k$) of several $\rm\AA$ width, which gives the 
local S/N for region $k$ as (S/N)$_{k} \equiv \langle c \rangle_{k}/ \sigma_{k}$.
We then averaged each (S/N)$_{k}$ to obtain S/N(measured) 
while weighting according to $w_{k} (\propto \sigma_{k}^{-2})$.
The correlation between S/N(predicted) and S/N(measured) is shown in figure 1, 
where we can see that both are reasonably correlated with each other (though 
quantitative agreement is not necessarily good). From a modest standpoint, we 
adopted (the rounded value of) the smaller one of these two S/N ratios for each star. 
Unfortunately, the finally achieved S/N ratios were not satisfactory. 
While moderate values of $\sim$~100 were realized for brighter giant stars, 
we could gain only several tens for the case of fainter turn-off stars 
of $V \gtsim 12.5$. Accordingly, since the spectrum quality is not sufficient
for reliable abundance determination, we must carefully check how much errors 
are involved in the resulting abundances, as will be done in section 5.
The observational data (observed date, exposure time, S/N) of these M~67 stars 
are summarized in table 1. 

Besides, we also observed Pollux (K0~III), Procyon (F5 IV--V), and Ganymede
(substitute for the Sun) as the reference stars, for which the spectra turned out 
to be of sufficiently high quality (S/N of more than several hundreds). 
Our spectra in the 7680--7820~$\rm\AA$ region are displayed in figure 3 for each 
of the 19 program stars.

\section{Stellar parameters}

Regarding the common properties of M~67 member stars, we adopted
[Fe/H] = +0.02 (metallicity), $E(B-V) = +0.04$ (color excess), 
and $(m-M)_{0} = 9.60$ (true distant modulus), following Sandquist (2004).
The effective temperature ($T_{\rm eff}$) was evaluated from 
dereddened $(B-V)_{0} \; [\equiv (B-V) - E(B-V)] $ by invoking
Alonso, Arribas, and Mart\'{\i}nez-Roger's (1996) equation (1) as well as 
Alonso, Arribas, and Mart\'{\i}nez-Roger's (1999) table 2
for 7 non-giants ($B-V < 0.9$) and 9 giants ($B-V > 0.9$), respectively.
The bolometric luminosity ($L$) was derived from the absolute magnitude
$M_{V} \; [\equiv V -(m-M)_{0} -3.1 E(B-V)]$ and the bolometric correction 
(B.C.) calibrated by Alonso, Arribas, and Mart\'{\i}nez-Roger (1995) 
as well as Alonso et al. (1999) for non-giants and giants, respectively.
As such, our M~67 targets are plotted on the $\log L$ vs. $\log T_{\rm eff}$   
diagram shown in figure 2b, where Demarque et al.'s (2004) Yale--Yonsei 
theoretical solar-metallicity evolutionary tracks (Y2\_tracks\_a0o2newOS.gz\footnote{
$\langle$http://www.astro.yale.edu/demarque/yystar.html$\rangle$.}) 
as well as the 4~Gyr isochrone (YYiso\_v2.tar.gz\footnote{
$\langle$http://csaweb.yonsei.ac.kr/\~{ }kim/yyiso.html$\rangle$.}) 
are also drawn.

The surface gravity ($\log g$) of each star was then calculated as
\begin{equation}
\log (g/g_{\odot}) =  \log (M/M_{\odot}) - \log (L/L_{\odot}) 
   + 4 \log (T_{\rm eff}/T_{\rm eff\odot})
\end{equation}
where those with subscript ``$\odot$'' are the solar values and 
we adopted $M = 1.3 M_{\odot}$ (cf. figure 1b).

Regarding the microturbulence ($v_{\rm t}$), Takeda et al.'s (2013)
empirical formula (cf. equation (1) and equation (2) therein) was adopted
for non-giants ($T_{\rm eff} > 5000$~K).
As to giants ($T_{\rm eff} < 5000$~K), we assumed a $\log g$-dependent relation
\begin{equation}
v_{\rm t} = 0.01 + 1.30 \log g - 0.31 (\log g)^{2}
\end{equation}
(where $v_{\rm t}$ is in km~s$^{-1}$ and $g$ is in cm~s$^{-2}$), 
which we found to well represent (within $\sim \pm$~0.1--0.2~km~s$^{-1}$) 
the $v_{\rm t}$ data derived by Takeda et al. (2008; cf. figure 1c therein) 
for giants in the parameter range of 4600~K~$< T_{\rm eff} < 4900$~K
and $2.3 < \log g < 3.3$.

Concerning the reference stars, we assigned ($T_{\rm eff}$, $\log g$, [Fe/H], 
and $v_{\rm t}$) as follows:
(4904~K, 2.84, +0.06, and 1.3~km~s$^{-1}$) for Pollux (Takeda et al. 2008),
(6612~K, 4.00, $-0.02$, and 2.0~km~s$^{-1}$) for Procyon (Takeda et al. 2005), 
and (5780~K, 4.44, 0.00, and 1.0~km~s$^{-1}$) for the Sun.
These finally adopted atmospheric parameters for each star are summarized 
in table 2. Besides, how the $\log g$ and $v_{\rm t}$ of our M~67 targets 
depend characteristically upon $T_{\rm eff}$ is displayed in figure 4.

Such assigned parameters are compared in table 3 with those 
adopted in previous investigations on oxygen abundances in M~67 
(cf. section 1), where five M~67 stars common to our sample were 
analyzed in four studies. We can see from this table that both are 
reasonably consistent with each other in terms of $T_{\rm eff}$ and 
$\log g$, since the differences are $\ltsim 50$~K and $\ltsim 0.1$~dex
in most cases, respectively (except that Pancino et al.'s spectroscopic 
$\log g$ for S~1010 is by $\sim 0.4$~dex higher while their photometric 
$\log g$ is in agreement with ours). Regarding $v_{\rm t}$ for giants,
Tautvai\u{s}ien\.{e} et al. (2000) used systematically higher values 
(1.7--1.8~km~s$^{-1}$) than that (1.3~km~s$^{-1}$) adopted by us, 
while Yong et al.'s (2005) $v_{\rm t}$ (1.34~km~s$^{-1}$ for S~1010) 
is consistent with ours.

\section{Abundance determinations}

\subsection{Model atmospheres and spectral line data}

The model atmosphere for each star to be used for abundance derivations
was constructed by three-dimensionally interpolating Kurucz's (1993) ATLAS9 
model grid in terms of $T_{\rm eff}$, $\log g$, and [Fe/H].
Regarding the atomic parameters of spectral lines, we basically invoked 
the extensive compilation by Kurucz and Bell (1995), while applying appropriate 
adjustments to $\log gf$ values when necessary 
(see footnote 5 and note in table 4). 
By using these atmospheric models, 
we simulated the theoretical spectra in the 7680--7820~$\rm\AA$ region 
while assuming the metallicity-scaled solar abundances for all elements 
(and convolved them with the Gaussian broadening function corresponding 
to $R \sim 12000$), which are overplotted in figure 3.\footnote{
In this computation, we basically adopted Kurucz and Bell's (1995) compilation
for the atomic data. 
We noticed, however, that their original $gf$ values were not necessarily 
appropriate for quite a few lines in this wavelength region, to which adequate 
adjustments must be applied in order to accomplish a satisfactory match between 
theory and observation. Accordingly, we estimated the necessary corrections
($\Delta \log gf$) for the following 17 lines by comparing the observed 
spectrum of Ganymede and the theoretical solar flux spectrum simulated with 
the standard solar abundances:
Ni~{\sc i} 7715.583 (by +1.0~dex), S~{\sc i} 7725.046 (by $-0.7$~dex),
Fe~{\sc i} 7742.685 (by +0.3~dex), Si~{\sc i} 7745.101 (by $-0.5$~dex),
Mg~{\sc i} 7746.345 (by $-1.3$~dex), Si~{\sc i} 7750.010 (neglected),
Fe~{\sc i} 7770.279 (by $-0.7$~dex), Ca~{\sc i} 7771.239 (neglected),
Fe~{\sc i} 7771.427 (by $-0.7$~dex), Fe~{\sc i} 7772.597 (by $-0.1$~dex),
Ti~{\sc i} 7773.904 (neglected), Fe~{\sc i} 7774.001 (by +1.5~dex),
Ca~{\sc i} 7775.496 (neglected), Ca~{\sc i} 7775.763 (neglected),
Fe~{\sc i} 7780.552 (by +2.3~dex; see also table 4),
Si~{\sc i} 7799.180 (neglected), and Si~{\sc i} 7799.996 (by +1.5~dex).
Besides, 6 additional CN lines in the 7770--7777~$\rm\AA$ region used by 
Takeda, Sadakane, and Kawanomoto (1998) were also included with 
Eriksson and Toft's (1979) $gf$ values.
}

\subsection{Spectrum-fitting analysis}

A three-step procedure based on the spectrum-fitting technique was adopted 
for oxygen abundance determination, the main theme of this study. This is 
because that directly measuring the equivalent widths of O~{\sc i} 7771--5 
lines is not easy because of their weakness (especially for giant stars) 
as well as appreciable noises due to insufficient quality of our spectra:\\
--- (i) First, we fitted the theoretical synthetic spectrum with the observed 
spectrum by finding the most optimal (LTE) abundance solutions of 
$A^{\rm L}$(O),$A^{\rm L}$(Fe), and  $A^{\rm L}$(Ni) 
(along with the macro-broadening width $v_{\rm M}$ and the radial velocity),
while applying the automatic fitting algorithm (Takeda 1995) 
to the 7770--7792~$\rm\AA$ region (comprising O~{\sc i} 7771--5, Fe~{\sc i} 7780,
and Ni~{\sc i} 7788 lines). The adopted atomic data of the relevant O~{\sc i},
Fe~{\sc i}, and Ni~{\sc i} lines are summarized in table 4, 
for which we invoked Kurucz and Bell's (1995) compilation.
However, only the $\log gf$ value ($-0.066$) for the Fe~{\sc i} line at 
7780.552~$\rm\AA$ was exceptionally taken from Kurucz and Peytremann's (1975) 
old database as done by Takeda and Sadakane (1997),
since the value ($-2.361$) given in Kurucz and Bell (1995) is too small
and apparently inappropriate (it would lead to an unrealistically large solar 
Fe abundance).
Since we could not arrive at any converged solution 
for S~1288 and S~1056 because lines are overwhelmed by noises,
we abandoned $A$(O) determinations for these two stars. 
The eventually accomplished fit for each spectrum is shown in figure 5,
and the results of $A^{\rm L}$(O),$A^{\rm L}$(Fe), and  $A^{\rm L}$(Ni) 
are given in table 2.\\
--- (ii) Second, the equivalent width ($EW_{7774}$) of the O~{\sc i} line 
at 7774.17~$\rm\AA$ (the middle line of the triplet) was {\it inversely} 
computed from such established solution of $A^{\rm L}$(O) by using the same 
model atmosphere.\footnote{Actually, this computation was done also for 
the other two lines of the triplet at 7771.94 and 7775.39~$\rm\AA$.
These $EW_{7771}$ and $EW_{7775}$ are well correlated with $EW_{7774}$ through
the following relations: 
$EW_{7771} = 0.1 + 1.18 EW_{7774} + 5.61 \times 10^{-4} EW_{7774}^2$ and
$EW_{7775} = 0.3 + 0.76 EW_{7774} + 7.70 \times 10^{-4} EW_{7774}^2$,
where $EW$'s are expressed in m$\rm\AA$.
Naturally, he inequality relation $EW_{7771} > EW_{7774} > EW_{7775}$
always holds between these three $EW$s, reflecting the difference of $\log gf$.
}
Similarly, $EW_{7780}$ and $EW_{7788}$ for Fe~{\sc i} 7780.55 and Ni~{\sc i} 7788.94
lines were derived from  $A^{\rm L}$(Fe) and $A^{\rm L}$(Ni).\\
--- (iii) Finally, regarding oxygen, the non-LTE correction for O~{\sc i} 7774.17 
($\Delta^{\rm N}_{7774}$) was evaluated from $EW_{7774}$ by following the procedure 
described in Takeda (2003), which was applied to $A^{\rm L}$(O) (derived in step (i)) 
to obtain $A^{\rm N}$(O) (non-LTE oxygen abundance) as
$A^{\rm N}$(O) = $A^{\rm L}$(O) + $\Delta^{\rm N}_{7774}$.\footnote{
We may regard the oxygen abundance derived from the O~{\sc i} 7774.17 
line (the middle component) practically equivalent to that corresponding 
to the O~{\sc i} 7771--5 triplet as a whole for the following reason.
The non-LTE abundance for the triplet (the mean of 
$A^{\rm N}_{7771}$, $A^{\rm N}_{7774}$, and $A^{\rm N}_{7775}$) is 
expressed as $\langle A^{\rm N} \rangle \equiv A^{\rm L} + 
(\Delta^{\rm N}_{7771} + \Delta^{\rm N}_{7774} + \Delta^{\rm N}_{7775})/3$,
where the inequality relation 
$|\Delta^{\rm N}_{7771}| > |\Delta^{\rm N}_{7774}| > |\Delta^{\rm N}_{7775}|$
generally holds for the extents of (negative) non-LTE correction according to
Takeda (2003) because of the difference in the component strengths 
($EW_{7771} > EW_{7774} > EW_{7775}$; cf. footnote 6).
We actually found the following equations regarding the $\Delta^{\rm N}$ values 
of each lines derived for our the program stars: 
$\Delta^{\rm N}_{7771} \simeq 1.10 \Delta^{\rm N}_{7774}$
and $\Delta^{\rm N}_{7775} \simeq 0.85 \Delta^{\rm N}_{7774}$.
Given this situation, the difference between $\Delta^{\rm N}_{7774}$
and $(\Delta^{\rm N}_{7774} + \Delta^{\rm N}_{7774} + \Delta^{\rm N}_{7775})/3$
turned out to be very small and negligible ($< 0.01$~dex for all stars), 
by which we can state that $A^{\rm N}_{7774}$ is essentially equivalent to 
$\langle A^{\rm N} \rangle$.
} 
Such obtained results of $\Delta^{\rm N}_{7774}$ and $A^{\rm N}$(O) are 
also presented in table 2.

\section{Error analysis}

In the present case where spectra of insufficient quality are used, 
the primary source of errors accompanied with our abundance results is 
due to noises. We estimated this kind of abundance error by two independent ways:
(1) Based on expected uncertainties in the equivalent widths, and 
(2) directly analyzing a number of mock spectra (with artificially added noises)
in the same spectrum-fitting method.  Besides, we also have to evaluate
errors due to ambiguities in the adopted atmospheric parameters ($T_{\rm eff}$,
$\log g$, and $v_{\rm t}$).

\subsection{Noise-related error \#1: based on equivalent widths}

Regarding the error in $EW$, we invoked the formula derived by Cayrel (1988)
\begin{equation}
\delta EW \simeq 1.6 (w \delta x)^{1/2} \epsilon,
\end{equation}
where $\delta x$ is the pixel size (0.21~$\rm\AA$), $w$ is the
full-width at half maximum (corresponding to $\simeq 25$~km~s$^{-1}$
or $R\simeq 12000$), and $\epsilon \equiv ({\rm S/N})^{-1}$.
We thus determined the abundances for each of the perturbed 
$EW_{+} (\equiv EW + \delta EW)$ and $EW_{-} (\equiv EW - \delta EW)$,
respectively, from which the differences from the standard $A$ were
derived as $\delta A_{+} (>0)$ and $\delta A_{-} (<0)$.

\subsection{Noise-related error \#2: based on artificial spectra}

We first calculated the reference spectrum corresponding to the standard 
solutions of $A_{0}$(O),$A_{0}$(Fe), $A_{0}$(Ni), and $v_{\rm M0}$
for each star (obtained in subsection 4.2). Then, randomly-generated noises of 
normal distribution (corresponding to the S/N ratio) were added to 
this standard spectrum, and this process was repeated 100 times. 
In this way, 100 artificial spectra were generated, as shown in figure 6.  
Next, we tried abundance determinations in exactly the same manner 
as described in subsection 4.2, and obtained $A_{i}$(O), $A_{i}$(Fe), 
and $A_{i}$(Ni) for each spectrum $i$ ($i$ = 1, 2, $\cdots$, 100),
The distributions of the resulting ${A_{i}}$(O) are illustrated in figure 7.
Then, from this ensemble of ${A_{i}}$, the average $\langle A \rangle$ 
and the standard deviation ($\sigma_{A}$) were computed. We can regard this 
$\sigma_{A}$ (representing the dispersion of ${A_{i}}$) as an estimate of 
abundance errors caused by noises.

\subsection{Effect of atmospheric parameters}

The uncertainties in $A$ due to errors of atmospheric parameters were evaluated 
by repeating the analysis on $EW$ while perturbing the standard values of atmospheric 
parameters interchangeably by $\pm$2\% in $T_{\rm eff}$ (typical differences
between various determinations; cf. Takeda et al. 2005, 2008), 
$\pm 0.1$~dex in $\log g$ (typical differences between spectroscopic $\log g$
and directly determined $\log g$ as adopted in this study; cf. Takeda et al. 
2005, 2008), and $\pm$20\% in $v_{\rm t}$ (typical scatter around the
analytical formula used by us). 
We call these six kinds of abundance variations as
$\delta_{T+}$, $\delta_{T-}$, $\delta_{g+}$, $\delta_{g-}$, 
$\delta_{v+}$, and $\delta_{v-}$, respectively.

\subsection{Trend of each error source}

These abundance errors derived in the last three subsections 
[($\delta A_{-}$, $\delta A_{+}$), ($-\sigma_{A}$, $+\sigma_{A}$),
($\delta_{T+}$, $\delta_{T-}$), ($\delta_{g+}$, $\delta_{g-}$), 
($\delta_{v+}$, and $\delta_{v-}$)] along with equivalent widths ($EW$)
are graphically shown as functions of $T_{\rm eff}$ in figures 8 (O), 
9 (Fe), and 10 (Ni). In figure 8, the non-LTE corrections 
($\Delta^{\rm N}_{7774}$) are also plotted for comparison. 

We can see from these figures the following trends:\\
--- The noise-related errors ($\delta A$ or $\sigma_{A}$) are quantitatively 
more significant than the parameter-related ones in the present case.\\
--- Reflecting the fact that they are of the same origin, $|\delta A|$ and 
$\sigma_{A}$ are almost of the same extent. though the former tends to be
somewhat larger than the latter.\\
--- Regarding the parameter-related errors, $\delta_{T}$'s are more or less 
appreciable (particularly for O in red giants), $\delta_{g}$'s are negligibly small,
and $\delta_{v}$'s are important only for Fe and Ni (especially in red giants).\\
--- The extents of (negative) non-LTE correction for O~{\sc i} 7774 are considerably 
$EW$-dependent, and range from $\sim 0.1$~dex (red giants) and $\sim$~0.4--0.5~dex
(turn-off stars).

For evaluating the total error budget, we adopt $\sigma_{A}$ (which is more 
preferable than $\delta A$ from the viewpoint of authenticity in the derivation 
procedure) for S/N-related error and  
$\delta_{Tgv} [\equiv (\delta_{T}^{2} + \delta_{g}^{2} + \delta_{v}^{2})^{1/2}]$
for parameter-related error (where $\delta_{T}$ is the average of
$|\delta_{T+}|$ and $|\delta_{T-}|$; and the same applies also for $\delta_{g}$ and $\delta_{v}$).
That is, we regard $\pm \sqrt{\sigma_{A}^{2} + \delta_{Tgv}^{2}}$ as the error bars
attached to the abundance results obtained in section 4 (cf. table 2). 
Our final abundances of O, Fe, and Ni (with such estimated error bars) 
are plotted against $T_{\rm eff}$ in figure 11, where the results for Pollux, 
Procyon and Sun are shown for comparison. 

\section{Discussion}

\subsection{Abundances of Fe and Ni}

We first review the abundances of Fe and Ni (iron-group elements representative 
of the metallicity) for M~67 stars shown in figure 11b and figure 11c.
Dividing all 16 stars into giant group ($T_{\rm eff} < 5000$~K; 9 stars) 
and non-giant group ($T_{\rm eff} > 5000$~K; 7 stars), we obtain 
$\langle A({\rm Fe})\rangle$(giant) = $7.57 (\pm 0.39)$, 
$\langle A({\rm Fe})\rangle$(non-giant) = $7.61 (\pm 0.58)$, and
$\langle A({\rm Fe})\rangle$(all) = $7.59 (\pm 0.44)$ 
as the averaged abundances of Fe for each group (values in parentheses 
after $\pm$ are the standard deviations).
Meanwhile, the results for Ni are 
$\langle A({\rm Ni})\rangle$(giant) = $6.40 (\pm 0.43)$, 
$\langle A({\rm Ni})\rangle$(non-giant) = $6.44 (\pm 0.55)$, and
$\langle A({\rm Ni})\rangle$(all) = $6.42 (\pm 0.47)$.
These results suggest that (i) $A$(Fe) and $A$(Ni) do not show
any systematic dependence upon $T_{\rm eff}$, and (ii) 
they are more or less consistent with those of the standard stars
(Pollux, Procyon, and Sun) within the dispersion,
which can also be recognized by eye-inspection of figure 11b and figure 11c,
even though appreciable scatters amounting to $\pm$~0.4--0.5~dex are seen.
Accordingly, we may state that the derived abundances of Fe and Ni for M~67 
stars do not show any systematic trend, which does not contradict 
the uniformity in the metallicity within this cluster. This, in turn, 
may imply that the atmospheric parameters 
which we assigned to each star are almost reasonable.

\subsection{O abundances in M~67: almost uniform but one exception}

Our non-LTE oxygen abundances, which are  plotted against 
$T_{\rm eff}$ in figure 11a, appear to show nearly the same trend
seen for Fe and Ni (figure 11b and figure 11c). That is, 
$A$(O) values do not show any systematic dependence upon $T_{\rm eff}$   
and are almost consistent with those of standard stars, which means
that almost the same argument as given in subsection 6.1 may
hold also for this case of oxygen. 
We notice, however, that only one star (S~1054) exhibits conspicuously 
low $A$(O) compared to other M~67 stars by $\sim 1$~dex. 
Excluding this  S~1054, we obtain 
$\langle A({\rm O})\rangle$(giant but S1054) = $8.55 (\pm 0.19)$, 
$\langle A({\rm O})\rangle$(non-giant) = $8.63 (\pm 0.46)$, and
$\langle A({\rm O})\rangle$(all but S1054) = $8.59 (\pm 0.33)$.\footnote{
That the abundance dispersion in non-giant group (0.46 dex) is appreciably 
larger than that in giant group (0.19 dex) is apparently due to the poor 
spectrum quality for the former group (note that we had to use data 
with S/N ratios as low as $\sim$~20--30; cf. table 1).}
Such derived mean abundance ($\sim 8.6$) is reasonably consistent with 
those of Pollux (8.43), Procyon (8.84), and Sun (8.60) within $\ltsim 0.2$~dex. 
Accordingly, we may conclude that atmospheric oxygen abundances of 
M~67 stars studied by us are almost uniform (with one exception) at 
the near-solar composition\footnote{
Regarding two stars (S~1288 and S~1056), for which $A$(O) could not 
be determined, their spectrum feature at the position of O~{\sc i} lines 
(severely affected by noises) does not contradict the normal oxygen composition,
as shown in figure 5, where we assumed [O/Fe]~=~0 as fixed for these two stars.}
irrespective of the evolutionary phase from unevolved turn-off stars through 
well-evolved red giants. This may be regarded as a significant consequence, 
given that the O~{\sc i} line strengths in our M~67 samples widely differ 
from each other (by a factor amounting up to $\sim 10$; cf. figure 8a).  

\subsection{On the nature of S~1054}

The outlier behavior of $A$(O) for S~1054 is curious.
Is this an unusual star compared to other M~67 members?  
While Sandquist (2004) classified this star as ``RGB binary?'',
its position on the HR diagram is not peculiar as compared to other
red giants, which makes it less likely that this star actually be 
an unresolved binary. This argument is substantiated by the fact 
that its heliocentric radial velocities ($V_{\rm r}^{\rm hel}$) at 20 
different dates over the time span of $\sim 5400$~days reported by 
Mathieu et al. (1986) are quite stable; i.e.,  
$\langle V_{\rm r}^{\rm hel} \rangle$ = 33.5~km~s$^{-1}$ with a
standard deviation of only 0.4~km~s$^{-1}$.
Besides, it is almost no doubt that S~1054 belongs to M~67 membership,
as this $\langle V_{\rm r}^{\rm hel} \rangle$ is reasonably consistent
with those of other members. Accordingly, we can not help considering 
that S~1054 is an ordinary red giant star of M~67. 
Then, how should we interpret its markedly low $A$(O) (7.37)
for this star compared with the others ($\sim 8.6$ on the average)?
Here, we should recall that the error accompanied with $A$(O) of 
S~1054 derived in subsection 5.4 is appreciably large to be  
$\pm 0.50 (\equiv \pm \sqrt{\sigma_{A}^{2} + \delta_{Tgv}^{2}})$
despite that spectrum S/N ratio of $\sim 100$ is sufficiently high, 
because the O~{\sc i} 7771--5 lines are considerably weak.
Therefore, the disagreement of $\sim 1.2$~dex corresponds to
$\sim 2.4 \sigma$, which means that a small probability of a few per cent\footnote{
In the case of normal (Gaussian) error distribution, the probability 
that a deviation of $\ge 2\sigma$ occurs is  0.046 (= 1 $-$ erf(2)). 
} still remains that this large discrepancy is due to a random fluctuation. 
In any case, this star is worth special attention, for which 
further study based on high-quality spectrum is eagerly awaited.
If this marked O-deficiency were real, this would make a very 
interesting object.

\subsection{Does evolution-induced mixing have impact on surface oxygen in red giants?}

Let us discuss the implication of this result from the viewpoint of our original 
motivation (i.e., to check the theoretical prediction in comparison with the observational
fact). According the recent calculation of Lagarde et al. (2012), the logarithmic oxygen
abundance ratio ([O/H]) of a solar-metallicity 1.25~$M_{\odot}$ star at the red-giant phase 
is only $-0.001$ (case 1: standard recipe) and $-0.0013$ (case 2: including thermohaline 
convection and rotation-induced mixing); i.e., essentially unaffected by the mixing 
of H-burning products for both of the cases considered by them.
In this sense, our observational consequence (no meaningful difference between the 
O abundances of evolved red giants and unevolved or less evolved stars) is fully
consistent with such theoretical expectations, as far as the present case of 
$\sim 1.3 M_{\odot}$ star is concerned.

However, little can be said about higher mass stars based on our result alone.
The apparent deficiency of O ($-0.5 \ltsim$~[O/Fe]~$\ltsim 0$) reported by 
Takeda et al. (2008; cf. Fig. 12 therein), which appears to be enhanced with mass and 
correlated with a deficit in C ($-0.5 \ltsim$~[C/Fe]~$\ltsim 0$) as well as an overabundance of 
Na ($0 \ltsim$~[C/Fe]~$\ltsim +0.4$), was actually derived from red giants of low- through 
intermediate-mass ($1~M_{\odot} \ltsim M \ltsim 5 M_{\odot}$).
While Lagarde et al.'s (2012) calculations for solar-metallicity stars suggest
a mass-dependent O-deficiency ([O/H] =  $-0.004$/$-0.02$, $-0.03$/$-0.04$, $-0.04$/$-0.05$,
$-0.07$/$-0.07$ corresponding to case~1/case~2 for 2, 3, 4, and 6~$M_{\odot}$,
respectively), its extent is marginal and evidently insufficient to account for
the trend suggested from Takeda et al.'s (2008) analysis based on the [O~{\sc i}] 5577 line. 
Whether or not the theoretical prediction reasonably explains the oxygen abundances 
also for red giants of higher-mass ($\sim$~2--5~$M_{\odot}$) remains yet to be investigated.

\subsection{Comparison with previous studies}

Finally, we review the results of $\langle$[O/H]$\rangle$ (mean of M~67 stars) 
obtained in this study in comparison with those published by others, 
which are summarized in table 5. We can recognize from this table that both of
our $\langle$[O/H]$\rangle$(giants) and $\langle$[O/H]$\rangle$(non-giants)
are $\sim 0.0$ and thus consistent with the previous results based on
high-quality data obtained by larger telescopes. Yet, we also note
that the apparent star-to-star dispersion (standard deviation) of our O abundances 
is considerably large (0.19~dex for giants excepting S~1054 and 0.46~dex for
non-giants) compared to literature values (mostly $< 0.1$~dex).
This is apparently due to large errors involved in our abundance analysis 
where considerably low-quality data had to be invoked, which stems from
the fact that our telescope and equipment were comparatively powerless 
from the viewpoint of the present-day standard. 

As mentioned in section 1, it was our alternative aim to check how much 
information can be gained from our rather noisy spectra of medium resolution.
We must admit here that the results we obtained are evidently insufficient 
for quantitatively discussing the detailed nature of abundances within 
the cluster (e.g., such as pursuing the degree of chemical homogeneity). 
We consider, however, that our study could serve as a useful pilot study
to clarify the general qualitative tendency (e.g., consistency between 
the giant and non-giant groups at near-solar oxygen abundances)
or to suggest an interesting candidate (S~1054) which may possibly have 
a peculiar oxygen composition. At any rate, we would stress the necessity
of verifying what has been implied in this study, based on new observational
data of much higher quality. 

\section{Conclusion}

Despite the considerable progress in the study of abundance changes in 
the surface of red giants caused by evolution-induced dredge-up of 
nuclear-processed material, no consensus has ever been accomplished yet 
about whether oxygen is significantly affected by such envelope mixing.

Give that the best approach to check if surface abundances
suffer appreciable changes on the way to the red giant phase is to investigate 
cluster stars formed with the same composition, we carried out a spectroscopic study
for selected 16 stars of M~67 in various evolutionary stages from the turn-off point 
through the red giant branch, along with Pollux, Procyon, and Sun (Ganymede)
as reference stars.

The observations were done by using the 2-m NAYUTA telescope and the 
MALLS spectrograph at Nishi-Harima Astronomical Observatory, by which 
we could obtain moderate-dispersion ($R\sim 12000$) spectra (centered around 
$\sim$~7700--7800~$\rm\AA$) for each star.

The atmospheric parameters were evaluated photometrically from colors ($T_{\rm eff}$), 
directly from mass and radius ($\log g$), and by using empirical formulas ($v_{\rm t}$).

We determined the oxygen abundances based on the synthetic spectrum-fitting 
in the 7770--7792~$\rm\AA$ region comprising O~{\sc i} 7771--5, Fe~{\sc i} 7780, 
and Ni~{\sc i} 7788 lines. The non-LTE correction was further taken into account 
for deriving the final O-abundances.

In estimating abundance ambiguities involved in our analysis, errors 
due to noises as well as due to uncertainties in the adopted atmospheric
parameters were taken into consideration. Regarding noise-related errors,
we tried two independent determinations (application of a conventional formula 
to $EW$ values and simulation by using many mock spectra with artificial noises)
and compared with each other. 

The abundances of Fe and Ni obtained as by-products were found to be roughly 
consistent with each other (though with a rather large dispersion)
without any systematic trend.

It turned out that the mean oxygen abundances of M~67 stars are quite similar between 
the giant ($T_{\rm eff} < 5000$~K) and non-giant ($T_{\rm eff} > 5000$~K) groups 
at the near-solar composition (with only one exception of S~1054 which shows 
a marked O-deficiency by $\sim 1$~dex).

This result implies that oxygen content in the stellar envelope is almost 
unaffected by the mixing of H-burning products in the red-giant phase, 
as far as M~67 stars of low mass ($\sim 1.3 M_{\odot}$) are concerned, which is 
in agreement with the prediction from the conventional stellar evolution theory 
of first dredge-up.

S~1054 is a puzzling object. Although we can not completely exclude the possibility 
that its considerably low O-abundance is spurious (i.e., simply due to random 
fluctuation), a more detailed reanalysis based on high-quality data is needed. 

The results we obtained based on rather noisy spectra of medium resolution 
(reflecting our use of smaller telescope) are not sufficient for quantitatively 
discussing the detailed nature of abundances within the cluster. 
Still, our study may be regarded as useful for a pilot study
to clarify the general qualitative behavior of O abundances in M~67.

\bigskip

This research has made use of the SIMBAD database (operated at CDS, Strasbourg, France) 
as well as the WEBDA database (operated at the Department of Theoretical Physics and 
Astrophysics of the Masaryk University).

\clearpage

\setcounter{table}{0}
\footnotesize
\begin{table}[h]
\caption{Basic data of observed M67 stars.}
\begin{center}
\begin{tabular}{crccccrrcl}\hline\hline
Star &  WEBDA &RA(2000) & DEC(2000) & $V$ & $B-V$ & $t_{\rm exp}$ & S/N & Date & Remark \\
     &  & (hour min sec)  & ($^{\circ}$~~~$'$~~~$''$) & (mag) & (mag) & (sec) &  & (2014,UT) & \\ 
\hline
S~1279 & 164 & 08 51 28.991 & +11 50 33.13 & 10.5379 & 1.1144 &  7200 & 130 & Jan04   & RGC \\
S~1010 & 141 & 08 51 22.804 & +11 48 01.78 & 10.4675 & 1.0946 &  7200 &  70 & Jan28   & RGC \\
S~1074 &  84 & 08 51 12.697 & +11 52 42.44 & 10.5221 & 1.0924 &  9600 & 120 & Jan29+30& RGC \\
S~1084 & 151 & 08 51 26.181 & +11 53 51.96 & 10.4929 & 1.0855 &  3600 &  75 & Feb01   & RGC \\
S~1054 & 104 & 08 51 17.028 & +11 50 46.40 & 11.1393 & 1.0810 &  4800 &  95 & Jan31   & RGB binary? \\
S~1288 & 217 & 08 51 42.363 & +11 51 23.09 & 11.2502 & 1.0714 &  9600 &  55 & Feb02+03& RGB binary? \\
S~989  & 135 & 08 51 21.565 & +11 46 06.20 & 11.4291 & 1.0702 &  8400 & 120 & Feb22   & Low GB \\
S~1277 & 218 & 08 51 42.314 & +11 50 07.82 & 11.6221 & 1.0367 &  4800 &  25 & Feb03   & Low GB \\
S~1293 &3035 & 08 51 39.386 & +11 51 45.66 & 12.1223 & 0.9952 &  7200 &  65 & Jan31   & RGB binary? \\
S~1245 & 227 & 08 51 44.741 & +11 46 46.03 & 12.9475 & 0.8925 &  6000 &  45 & Jan31   & SG \\
S~1056 &  96 & 08 51 15.639 & +11 50 56.22 & 13.0015 & 0.8379 &  7200 &  20 & Jan27   & SG \\
S~489  &7489 & 08 50 16.296 & +11 53 48.00 & 12.7600 & 0.6800 & 13200 &  25 & Jan28   & TO gap \\
S~1034 & 115 & 08 51 18.541 & +11 49 21.52 & 12.6530 & 0.6056 & 14400 &  25 & Jan04   & SG \\
S~1083 & 157 & 08 51 27.423 & +11 53 26.56 & 12.7547 & 0.5678 &  6000 &  50 & Jan27   & TO \\
S~1270 & 241 & 08 51 49.143 & +11 49 43.59 & 12.6843 & 0.5578 & 10800 &  40 & Jan31   & SG \\
S~1456 & 255 & 08 51 53.354 & +11 48 20.89 & 12.7160 & 0.5181 & 10800 &  25 & Jan28   & TO \\
\hline
\end{tabular}
\end{center}
\footnotesize
Note.\\
Most columns are self-explanatory. 
Each M~67 star, designated with the serial number assigned by Sanders (1977)
and the reference number of WEBDA database (Mermilliod 1995; available at 
$\langle$http://www.univie.ac.at/webda/$\rangle$),
is arranged as in table~2 (decreasing order in $B-V$, 
or increasing order in $T_{\rm eff}$).
The positional data are taken from SIMBAD database, while the photometric data
($V$, $B-V$) are from Sandquist (2004). 
See section 2 for the derivation of S/N ratios presented in column 8.
Given in column 10 is the evolutionary status classified by Sandquist (2004). 

\end{table}

\newpage 
\setcounter{table}{1}
\setlength{\tabcolsep}{3pt}
\begin{table}[h]
\small
\caption{Stellar parameters and the resulting abundances.} 
\begin{center}
\begin{tabular}
{ccccc cccr  cr  cr} 
\hline \hline
Star & $T_{\rm eff}$ & $\log g$ & [Fe/H] & $v_{\rm t}$ &
$A^{\rm L}_{\rm O}$ & $\Delta^{\rm N}_{\rm O}$ & $A^{\rm N}_{\rm O}$ & $EW_{\rm O}$ & 
$A^{\rm L}_{\rm Fe}$ & $EW_{\rm Fe}$ & $A^{\rm L}_{\rm Ni}$ & $EW_{\rm Ni}$ \\
\hline
Pollux  & 4904& 2.84& +0.06& 1.3&  8.53& $-$0.10&  8.43&  25&  7.47& 135&  6.56& 133 \\
\hline
S~1279   & 4668& 2.45& +0.02& 1.3&  8.58& $-$0.09&  8.49&  20&  7.45& 140&  5.94& 109 \\
S~1010   & 4705& 2.44& +0.02& 1.3&  8.57& $-$0.09&  8.48&  21&  6.90& 108&  5.64&  91 \\
S~1074   & 4709& 2.47& +0.02& 1.3&  8.74& $-$0.10&  8.64&  25&  7.11& 119&  6.08& 116 \\
S~1084   & 4722& 2.46& +0.02& 1.3&  8.34& $-$0.08&  8.26&  16&  7.78& 162&  6.95& 172 \\
S~1054   & 4731& 2.73& +0.02& 1.3&  7.43& $-$0.06&  7.37&   3&  7.78& 162&  6.77& 154 \\
S~1288   & 4749& 2.78& +0.02& 1.2& $\cdots$ & $\cdots$ & $\cdots$ & $\cdots$ &  8.20& 200&  6.39& 124 \\
S~0989   & 4751& 2.85& +0.02& 1.2&  8.53& $-$0.08&  8.45&  18&  7.75& 154&  6.73& 144 \\
S~1277   & 4816& 2.96& +0.02& 1.2&  8.97& $-$0.10&  8.87&  30&  7.70& 149&  6.61& 134 \\
S~1293   & 4900& 3.21& +0.02& 1.0&  8.74& $-$0.08&  8.66&  26&  7.49& 125&  6.47& 112 \\
S~1245   & 5044& 3.58& +0.02& 0.9&  8.64& $-$0.07&  8.57&  25&  7.99& 165&  6.10&  82 \\
S~1056   & 5199& 3.68& +0.02& 1.0& $\cdots$ & $\cdots$ & $\cdots$ & $\cdots$ & $\cdots$ & $\cdots$ &  7.22& 147 \\
S~0489   & 5703& 3.79& +0.02& 1.2&  8.72& $-$0.15&  8.57&  57& $\cdots$ & $\cdots$ &  5.93&  54 \\
S~1034   & 5975& 3.84& +0.02& 1.4&  9.49& $-$0.35&  9.14& 128&  7.07&  80&  6.80&  90 \\
S~1083   & 6123& 3.93& +0.02& 1.5&  8.45& $-$0.16&  8.29&  63&  6.89&  67&  5.80&  28 \\
S~1270   & 6163& 3.92& +0.02& 1.6&  9.62& $-$0.44&  9.18& 155&  7.96& 130&  6.96&  94 \\
S~1456   & 6329& 3.98& +0.02& 1.7&  8.19& $-$0.16&  8.03&  56&  8.13& 137&  6.29&  45 \\
\hline
Procyon & 6612& 4.00& $-$0.02& 2.0&  9.21& $-$0.37&  8.84& 148&  7.38&  87&  6.53&  47 \\
Sun& 5780& 4.44&  0.00& 1.0&  8.69& $-$0.09&  8.60&  48&  7.74& 124&  6.61&  81 \\
\hline
\end{tabular}
\end{center}
Note. \\
In columns 1 through 5 are given the name (M~67 stars are indicated by 
Sanders numbers), effective temperature (in K), logarithmic surface gravity 
(in cm~s$^{-2}$), metallicity (logarithmic Fe abundance relative to the Sun; in dex), 
and microturbulent velocity dispersion (in km~s$^{-1}$) for each star. 
The LTE abundances ($A^{\rm L}$) resulting from our spectrum fitting analysis and 
the corresponding equivalent widths ($EW$; in m$\rm\AA$) evaluated for O~{\sc i} 7774.17, 
Fe~{\sc i} 7780.55, and Ni~{\sc i} 7788.94 lines are presented in the remaining columns, 
where the non-LTE corrections ($\Delta^{\rm N}_{\rm O}$) and the non-LTE abundances 
($A^{\rm N}_{\rm O}$) are also given regarding oxygen. All abundance results are expressed 
in the usual normalization of $A_{\rm H}$~=~12.00.
\end{table}

\setcounter{table}{2}
\begin{table}[h]
\caption{Comparison of the atmospheric parameters with the literature values.}
\small
\begin{center}
\begin{tabular}{c ccc l}\hline\hline
Star & $T_{\rm eff}$ & $\log g$ & $v_{\rm t}$ & References \\
  & (K) & (cm~s$^{-1}$) & (km~s$^{-1}$)  &   \\
\hline
S~1279 (F164)& 4700 & 2.50 & 1.8 & Tautvai\u{s}ien\.{e} et al. (2000) \\
              & 4668 & 2.45 & 1.3 & This study \\
\hline
S~1010 (F141)& 4730 & 2.40 & 1.8 & Tautvai\u{s}ien\.{e} et al. (2000) \\
              & 4700 & 2.30 & 1.34& Yong et al. (2005) (spectroscopic)\\
              & 4640 & 2.30 &     & Yong et al. (2005) (photometric)\\
              & 4650 & 2.80 & 1.3 & Pancino et al. (2010) (spectroscopic)\\
              & 4590 & 2.42 & 1.2/1.8 & Pancino et al. (2010) (photometric)\\
              & 4705 & 2.44 & 1.3 & This study \\
\hline
S~1074 (F84) & 4750 & 2.40 & 1.8 & Tautvai\u{s}ien\.{e} et al. (2000) \\
              & 4709 & 2.47 & 1.3 & This study \\
\hline
S~1084 (F151)& 4760 & 2.40 & 1.7 & Tautvai\u{s}ien\.{e} et al. (2000) \\
              & 4722 & 2.46 & 1.3 & This study \\
\hline
S~1034        & 6019 & 4.00 & 1.5 & Randich et al. (2006) \\
              & 5975 & 3.84 & 1.4 & This study \\
\hline
\end{tabular}
\end{center}
\end{table}

\clearpage
\setcounter{table}{3}
\footnotesize
\begin{table}[h]
\caption{Atomic data of the spectral lines used for abundance determinations.}
\begin{center}
\begin{tabular}{ccccccc}\hline\hline
Species & $\lambda_{\rm air}$ & $\chi_{\rm low}$ & $\log gf$ & Gammar & Gammas &
Gammaw \\
        & ($\rm\AA$) & (eV) & (dex) & (dex) & (dex) & (dex) \\
\hline
O~{\sc i} & 7771.944 & 9.146&+0.324 & 7.52 & $-$5.55 & ($-$7.65) \\
O~{\sc i} & 7774.166 & 9.146&+0.174 & 7.52 & $-$5.55 & ($-$7.65) \\
O~{\sc i} & 7775.388 & 9.146&$-$0.046 & 7.52 & $-$5.55 & ($-$7.65) \\
Fe~{\sc i}& 7780.552 & 4.473&$-$0.066 & 7.88 & $-$6.13 &$-$7.80 \\
Ni~{\sc i}& 7788.936 & 1.951&$-$2.420 & 8.00 & $-$6.31 &$-$7.85 \\
\hline
\end{tabular}
\end{center}
\footnotesize
Note.\\
The first four columns are self-explanatory.
Presented in columns 5--7 are the damping parameters: 
Gammar is the radiation damping width (s$^{-1}$) [$\log\gamma_{\rm rad}$], 
Gammas is the Stark damping width (s$^{-1}$) per electron density (cm$^{-3}$) 
at $10^{4}$ K [$\log(\gamma_{\rm e}/N_{\rm e})$], and
Gammaw is the van der Waals damping width (s$^{-1}$) per hydrogen density 
(cm$^{-3}$) at $10^{4}$ K [$\log(\gamma_{\rm w}/N_{\rm H})$]. \\
Most of these data were taken from the compilation of Kurucz and Bell (1995),
though Kurucz and Peytremann's (1975) $\log gf$ value was adopted for Fe~{\sc i} 7780.552 
as done in Takeda and Sadakane (1997) (cf. subsection 4.2).
The parenthesized damping parameters are the default values computed by the WIDTH9 program, 
since the data are not available in Kurucz and Bell (1995).
\end{table}

\clearpage
\setcounter{table}{4}
\begin{table}[h]
\caption{Mean oxygen abundances of M~67 stars derived by previous studies.}
\footnotesize
\begin{center}
\begin{tabular}{cccccccccl}\hline\hline
Ref. & Telescope & Type & N & Line & $EW$ & S/N & $R$ & $\langle$[O/H]$\rangle$ $(\pm \sigma)$ & Remark \\ 
(1)  &   (2)     & (3)  & (4) & (5) & (6) & (7) & (8) & (9)  & (10) \\
\hline
BRO85  & 2.7~m McDonald &  RG & 3 &  6300  & 26--39 & 70--100 & 25000    &    8.70$^{*}$~$(\pm 0.04)$ & Solar abundance unavailable\\
SHE00 & 2.7~m McDonald &  BS & 5 &  7773 & 84--296& 44--75  & 30000    & $+0.11$~$(\pm 0.31)$ & NLTE \\ 
SHE00 & 2.7~m McDonald &  TO & 4 &  7773 & 83--138& 35--69  & 30000    & $+0.04$~$(\pm 0.08)$ & NLTE \\ 
TAU00  & 2.6~m Nord. Opt. Tel. &  RG & 9 &  6300  &  3--6  & $\gtsim 100$& 30000 & $-0.01$~$(\pm 0.05)$ & $R\sim 60000$ data partly used \\
YON05  & 4.0~m KPNO/CTIO&  RG & 3 &  6300/6363 & N/A& 100--200& 28000    & $+0.09$~$(\pm 0.05)$ &  \\
RAN06  & 8.2~m VLT      &  TO & 10&  6300   &  5--10 & 90--180 & 45000    & $+0.02$~$(\pm 0.03)$ &  \\
PAC08  & 8.2~m VLT      &  ST & 5 &  6300   &  4--5 & $\sim 80$ & 100000  & $-0.04$~$(\pm 0.08)$ &  \\
PAN10  & 2.2~m Calar Alto & RG & 3 & 6300 & N/A & 70--100 & 30000 & $+0.09$~$(\pm 0.10)$ & Spectrum synthesis \\
\hline
TAKHON   & 2.0~m NAYUTA   &  RG & 7 &  7773 & 16--30&  25--130 & 12000    & $-0.05$~$(\pm 0.19)$ & S~1054 excluded, NLTE \\
TAKHON   & 2.0~m NAYUTA   & TO/SG & 6 & 7773 & 25--155&  25--50 & 12000    & $+0.03$~$(\pm 0.46)$ & NLTE \\
\hline
\end{tabular}
\end{center}
\footnotesize
Columns (1) through (10) gives the reference, used telescope, type of stars, number of stars,
adopted line (``6300'', ``6363'', ``7773'', means [O~{\sc i}] 6300,  [O~{\sc i}] 6363, and
O~{\sc i} 7771--5, respectively),
equivalent-width range (in m$\rm\AA$), S/N ratio range, resolving power, mean oxygen abundance
relative to the reference solar abundance along with the standard deviation (in dex), and remarks 
(if any), respectively.
Abbreviations used in columns (1) and (3) are denoted as follows:\\  
(1): BRO85 $\cdots$ Brown (1985); SHE00 $\cdots$ Shetrone and Sandquist (2000); 
TAU00 $\cdots$ Tautvai\u{s}ien\.{e} et al. (2000); YON05 $\cdots$ Yong et al. (2005);
RAN06 $\cdots$ Randich et al. (2006); PAC08 $\cdots$ Pace et al. (2008); 
PAN10 $\cdots$ Pancino et al. (2010); TAKHON $\cdots$ this study.\\
(3): RG $\cdots$ red giants; BS $\cdots$ blue stragglers; TO $\cdots$ turn-off stars;
ST $\cdots$ solar-type stars; SG $\cdots$ subgiants.\\
$^{*}$ $\langle A$(O)$\rangle$ (mean oxygen abundance) is presented here,
since the reference solar abundance is unavailable.
\end{table}

\newpage

\setcounter{figure}{0}
\begin{figure}
\begin{center}
  \includegraphics[width=8cm]{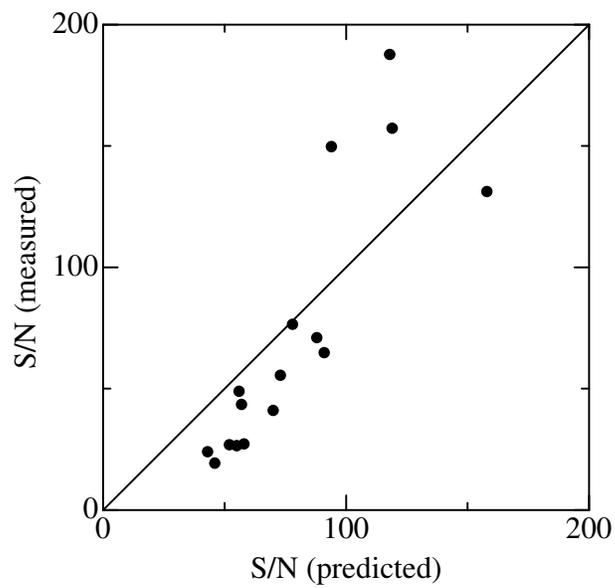}
\end{center}
\caption{
Correlation between S/N(predicted) (evaluated from photon counts)
and S/N (measured) (directly estimated from fluctuations 
at line-free windows) for 16 stars of M~67.
}
\end{figure}

\setcounter{figure}{1}
\begin{figure}
\begin{center}
  \FigureFile(80mm,80mm){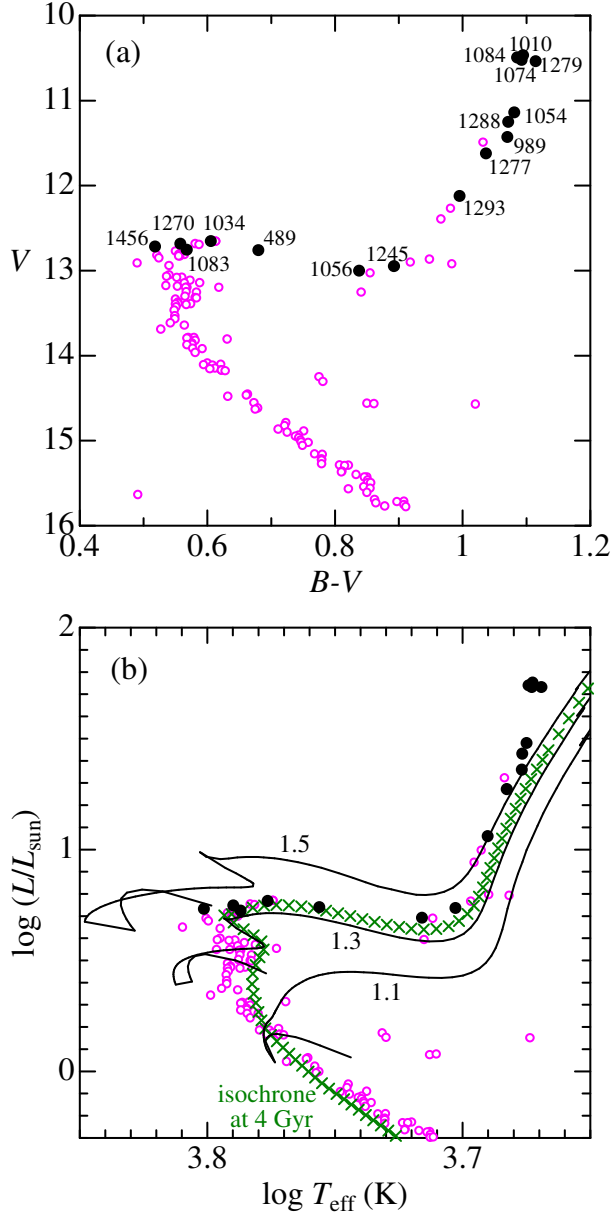}
\end{center}
\caption{
(a) M~67 stars plotted on $V$ vs. $B-V$ diagram, based on
the photometric data taken from Sandquist (2004). 
Filled circles $\cdots$ our 16 targets (from the turn-off 
point through the red-giant branch).
Open circles $\cdots$ other M~67 member stars which we did not observe.
(b) M~67 stars plotted on the $\log L$ vs. $\log T_{\rm eff}$ diagram,
with the same meanings of symbols as in panel (a). (See section 3 
for the derivation of $T_{\rm eff}$ and $L$ for each star.)
The Yale--Yonsei theoretical solar-metallicity evolutionary tracks 
(corresponding to 1.1, 1.3, and 1.5~$M_{\odot}$) as well as the 4~Gyr 
isochrone (Demarque et al. 2004) are also depicted by solid lines and 
crosses, respectively. 
}
\end{figure}

\setcounter{figure}{2}
\begin{figure}
\begin{center}
  \FigureFile(140mm,200mm){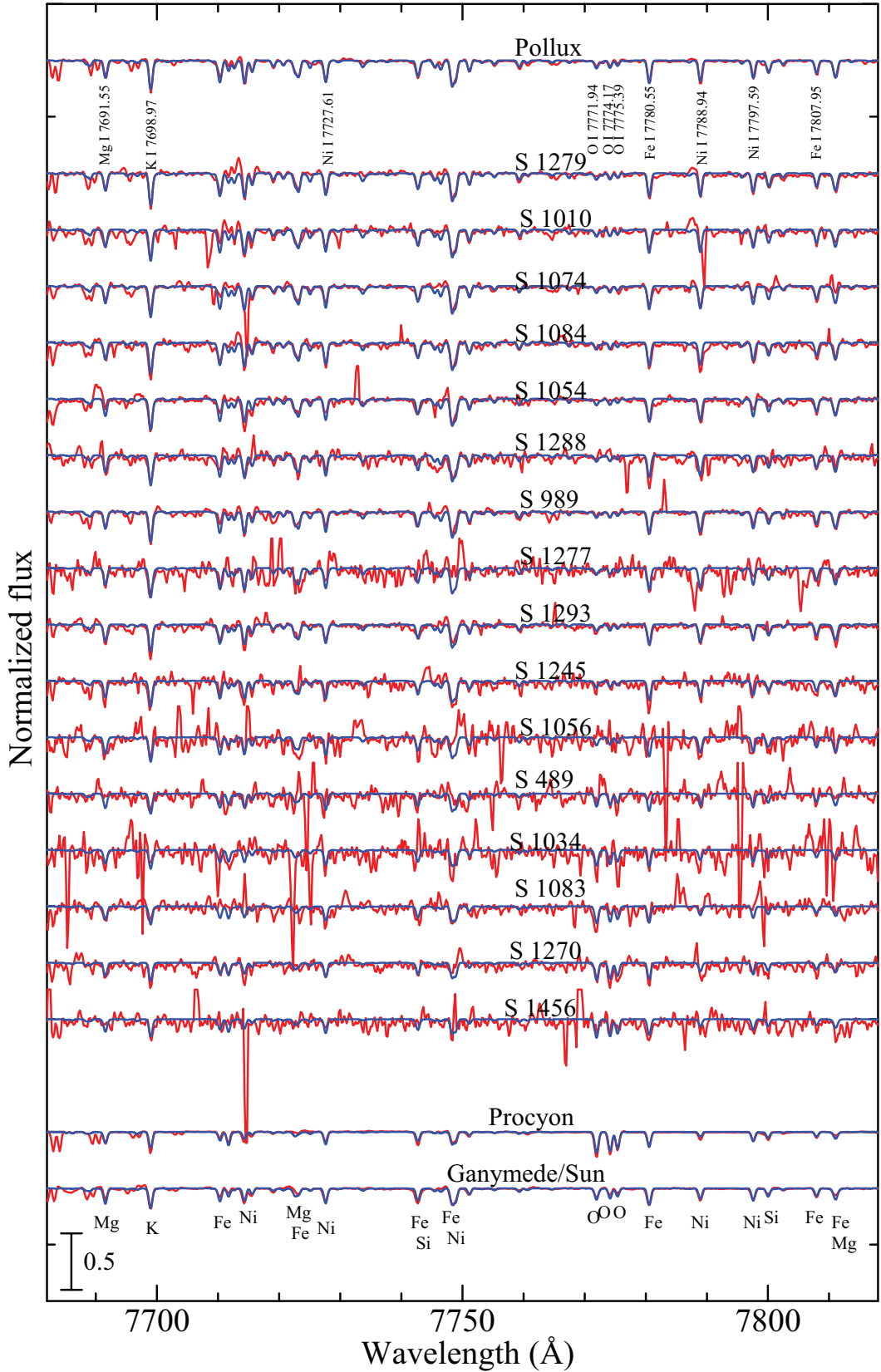}
\end{center}
\caption{
Depicted in red lines are the observed spectra of 16 targets in M~67 
(arranged in the order of increasing $T_{\rm eff}$ as in table 1),
where theoretical spectra simulated with scaled solar abundances 
(see subsection 4.1) are overplotted in blue lines. The spectra of Pollux, 
Procyon, and Ganymede (substitute for the Sun) are also shown for comparison. 
The wavelength scale of each spectrum is adjusted to the laboratory system.
}
\end{figure}

\setcounter{figure}{3}
\begin{figure}
\begin{center}
  \FigureFile(80mm,150mm){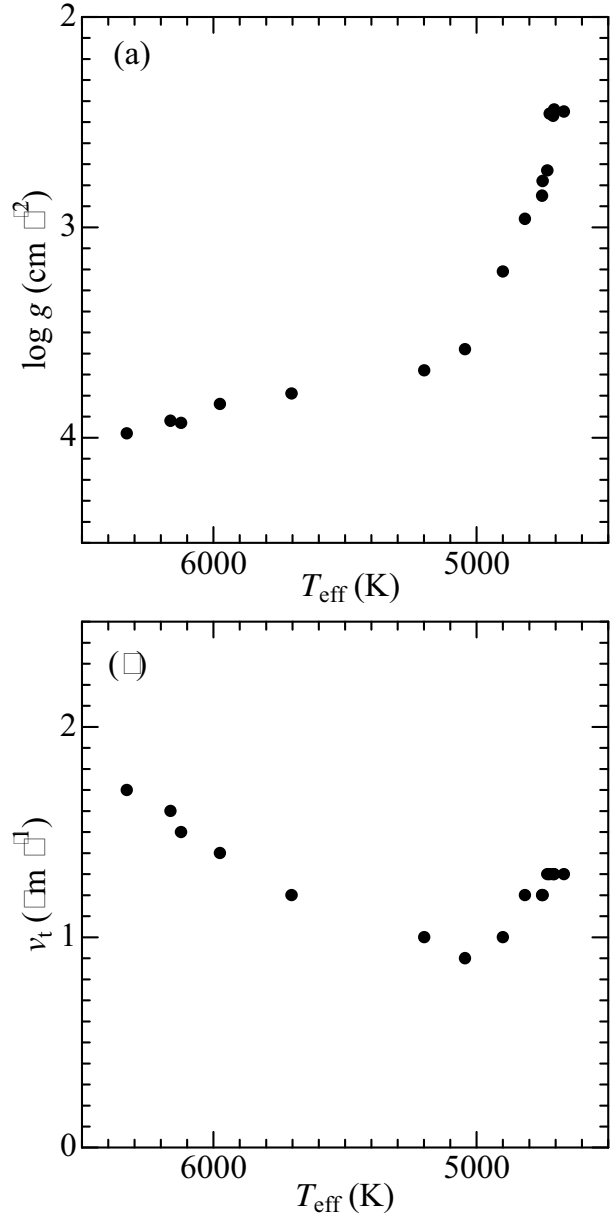}
\end{center}
\caption{
$T_{\rm eff}$-dependence of (a) $\log g$ and (b) $v_{\rm t}$
for our 16 targets of M~67. 
}
\end{figure}

\setcounter{figure}{4}
\begin{figure}
\begin{center}
  \FigureFile(140mm,200mm){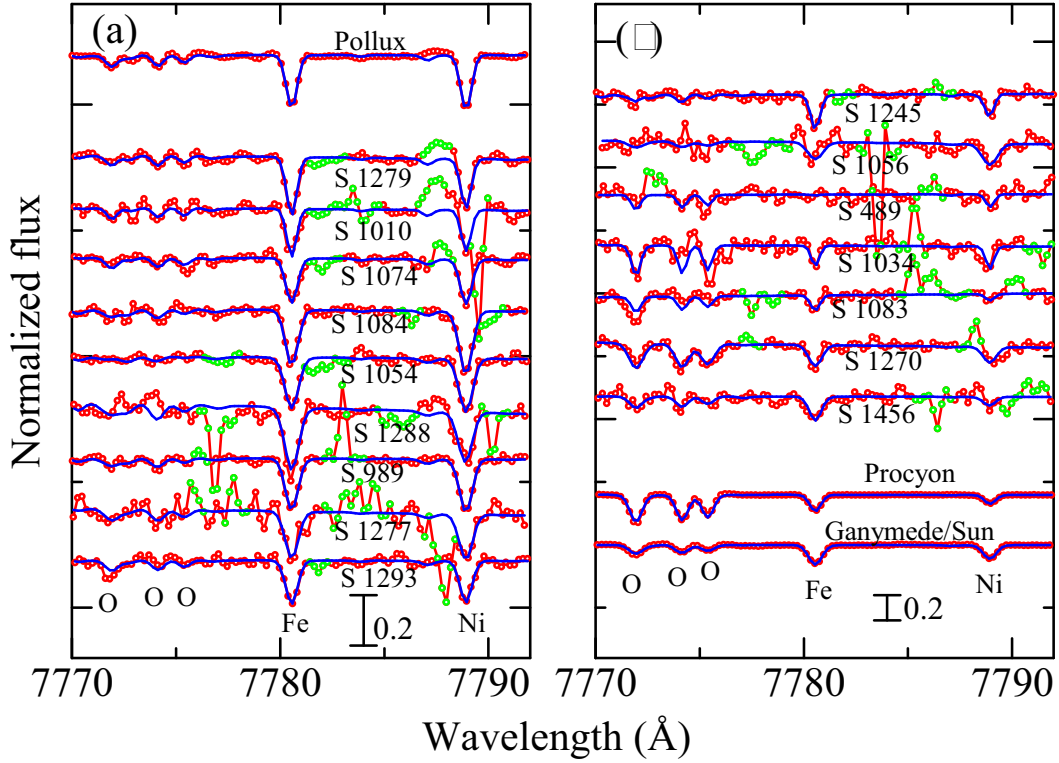}
\end{center}
\caption{
Synthetic spectrum fitting in the 7770--7792~$\rm\AA$ region 
comprising O~{\sc i} 7771.94, 7774.17, 7775.39, Fe~{\sc i} 7780.55, 
and Ni~{\sc i} 7788.94 lines.
The best-fit theoretical spectra are shown by blue solid lines. 
The observed data are plotted by red symbols connected by lines, 
where those rejected in the fitting are highlighted in green. 
The spectra of M~67 stars are arranged (from top to bottom) 
in the order of increasing $T_{\rm eff}$, Note that the ordinate 
scale is twice as different in the left panel (giant stars) 
as in the right panel (non-giant stars). 
For the $A$(O)-indeterminable cases of S~1288 and S~1056, the theoretical 
spectra shown here were simulated with the fixed oxygen abundances 
at the metallicity-scaled solar composition. 
}
\end{figure}

\setcounter{figure}{5}
\begin{figure}
\begin{center}
  \FigureFile(140mm,140mm){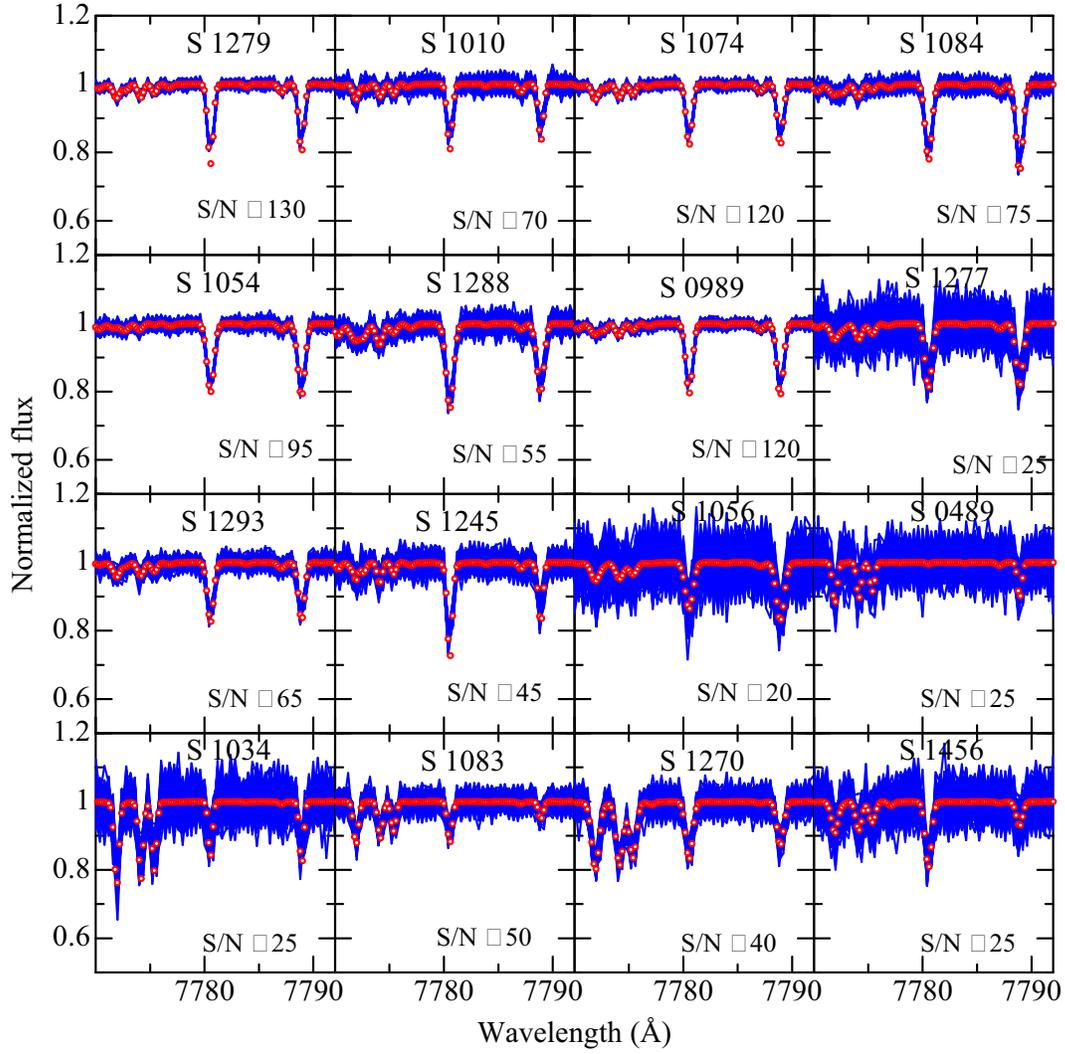}
\end{center}
\caption{
Modeled spectra in the 7770--7792~$\rm\AA$ region with artificial noises, for which
 spectrum-fitting analysis was formally performed to estimate abundance errors.
Blue solid lines are overplotted mock spectra (100 for each star), generated by adding 
random noises (relevant for the S/N ratio) to the theoretical spectrum (red open circles)
inversely computed based on the standard solutions of $A$(O), $A$(Fe), $A$(Ni), 
and $v_{\rm M}$.
}
\end{figure}

\setcounter{figure}{6}
\begin{figure}
\begin{center}
  \FigureFile(140mm,140mm){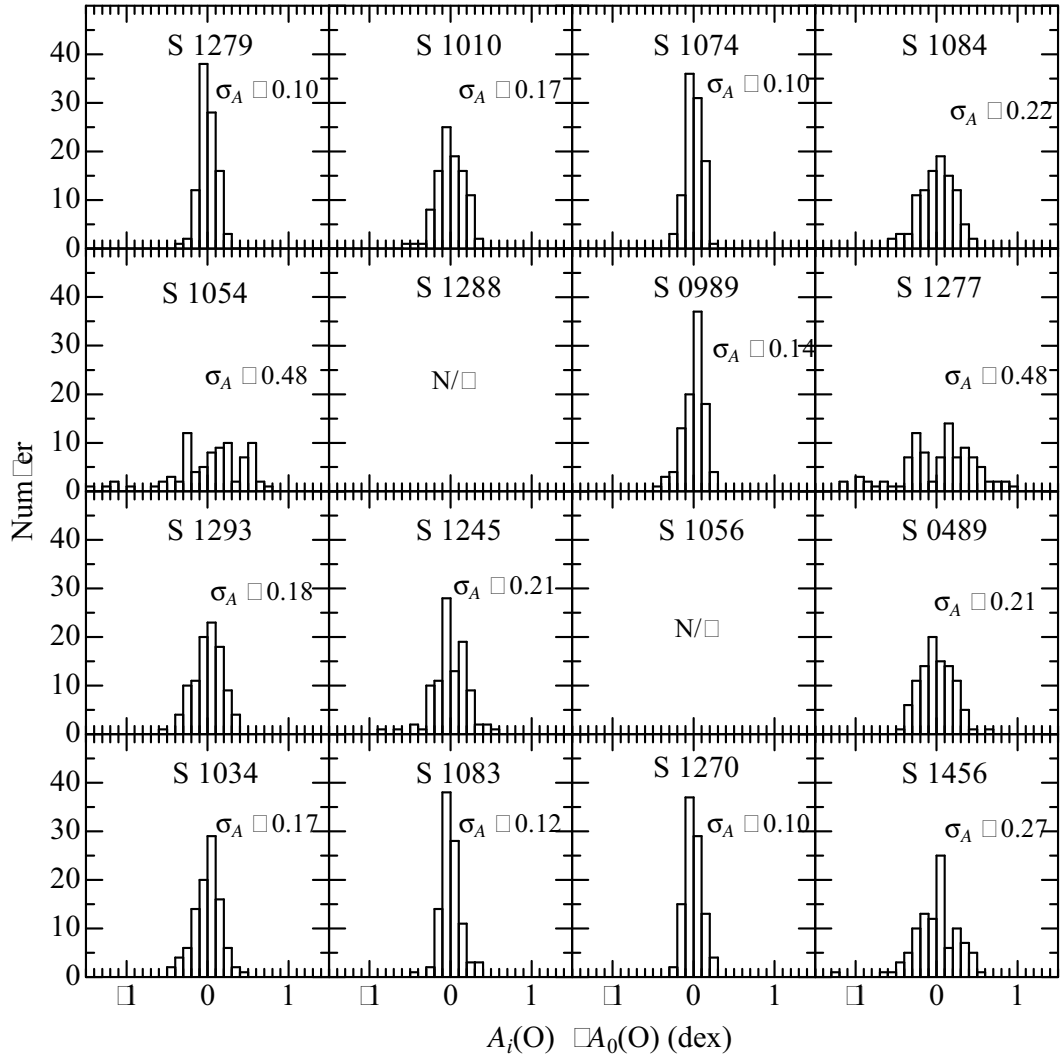}
\end{center}
\caption{
Histogram of oxygen abundance deviations, which are defined as the differences 
between the apparent $A_{i}$(O) solutions (resulting from the fitting-analysis of 
100 mock spectra with artificial noises; cf. figure 6) and the given $A_{0}$(O) 
used for modeling of the standard spectrum.
}
\end{figure}

\setcounter{figure}{7}
\begin{figure}
\begin{center}
  \FigureFile(100mm,140mm){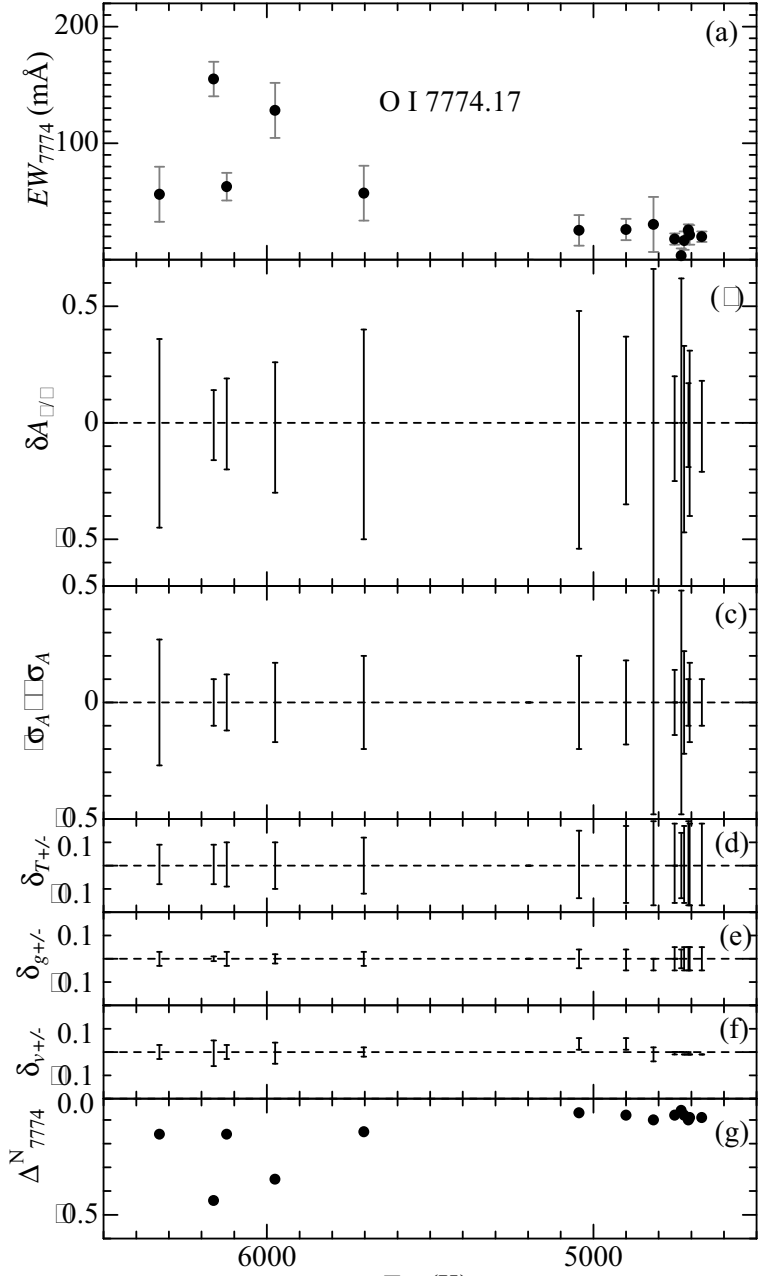}
\end{center}
\caption{
Quantities related to O abundance errors plotted against $T_{\rm eff}$. 
(a) $EW^{\rm O}_{7774}$ (equivalent width of O~{\sc i} 7774.17) with
$EW$ ambiguity of $\pm\delta EW$ as error bars, 
(b) $\delta A_{+}$ and $\delta A_{-}$ (abundance change corresponding to
perturbation of $\pm\delta EW$), 
(c) standard deviation ($\pm \sigma$) of abundance differences resulting 
from the experiment based on 100 mock spectra with artificial noises,
(d) $\delta_{T+}$ and $\delta_{T-}$ (abundance variations 
in response to $T_{\rm eff}$ changes of +2\% and $-2$\%), 
(e) $\delta_{g+}$ and $\delta_{g-}$ (abundance variations 
in response to $\log g$ changes of $+0.1$~dex and $-0.1$~dex), 
(f) $\delta_{v +}$ and $\delta_{v -}$ (abundance 
variations in response to perturbing $v_{\rm t}$
by +20\% and $-20$\%), and
(g) $\Delta^{\rm O}_{7774}$ (non-LTE correction for O~{\sc i} 7774.17).
Note that the same ordinate scales are adopted except for panel (a).
The signs of $\delta$'s are $\delta_{T+}<0$, $\delta_{T-}>0$,
$\delta_{g+}>0$, $\delta_{g-}<0$, $\delta_{v +}<0$, and
$\delta_{v -}>0$.
}
\end{figure}

\setcounter{figure}{8}
\begin{figure}
\begin{center}
  \FigureFile(100mm,140mm){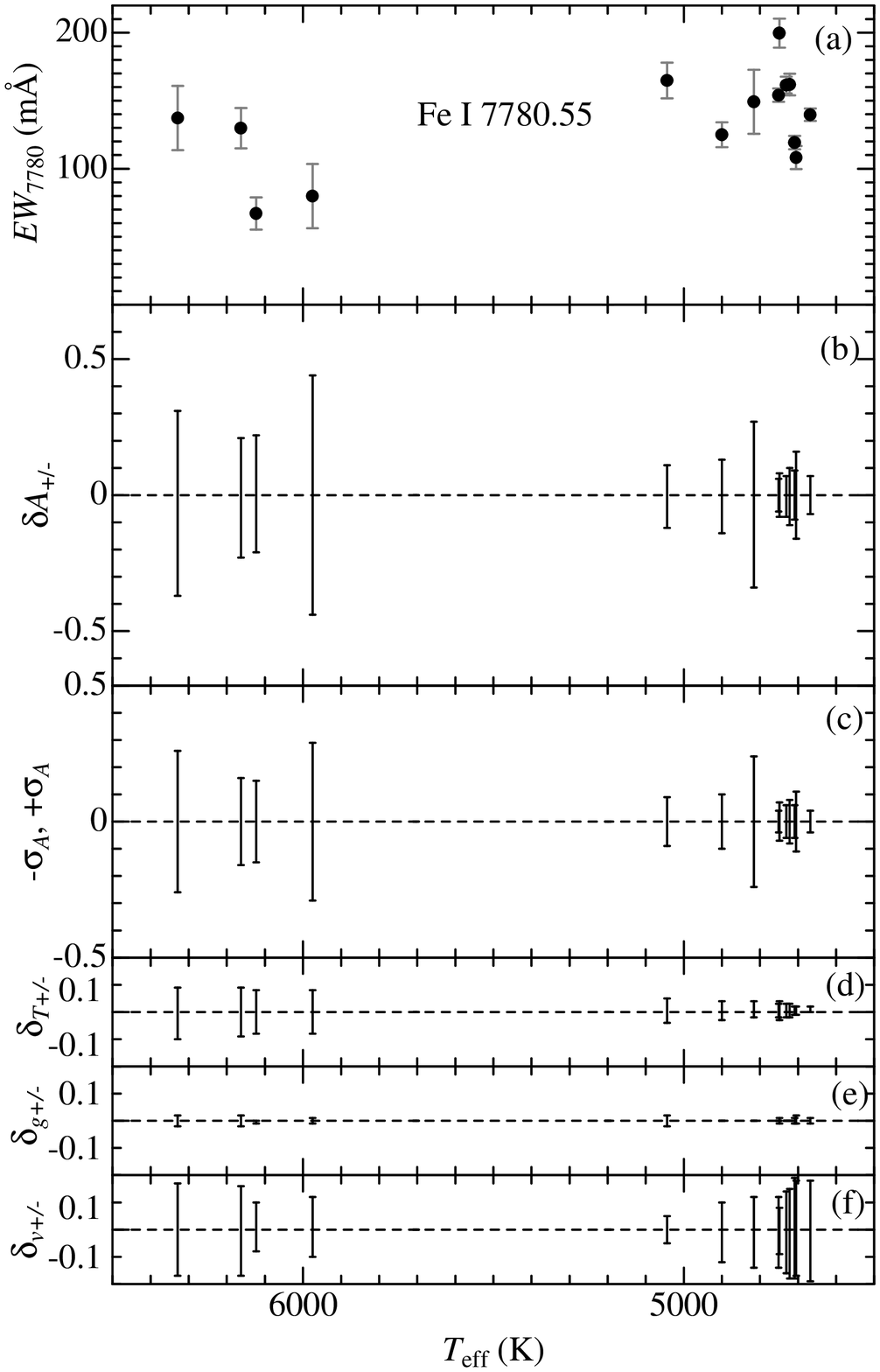}
\end{center}
\caption{
Quantities related to Fe abundance errors plotted against $T_{\rm eff}$. 
While panel (a) shows $EW^{\rm Fe}_{7780}$ (equivalent width of 
Fe~{\sc i} 7780.55), the meanings of panels (b)--(f) are the same as 
in figure 8. The signs of $\delta$'s are $\delta_{T+}>0$, $\delta_{T-}<0$,
$\delta_{g+}>0$, $\delta_{g-}<0$, $\delta_{v +}<0$, and
$\delta_{v -}>0$.
}
\end{figure}

\setcounter{figure}{9}
\begin{figure}
\begin{center}
  \FigureFile(100mm,140mm){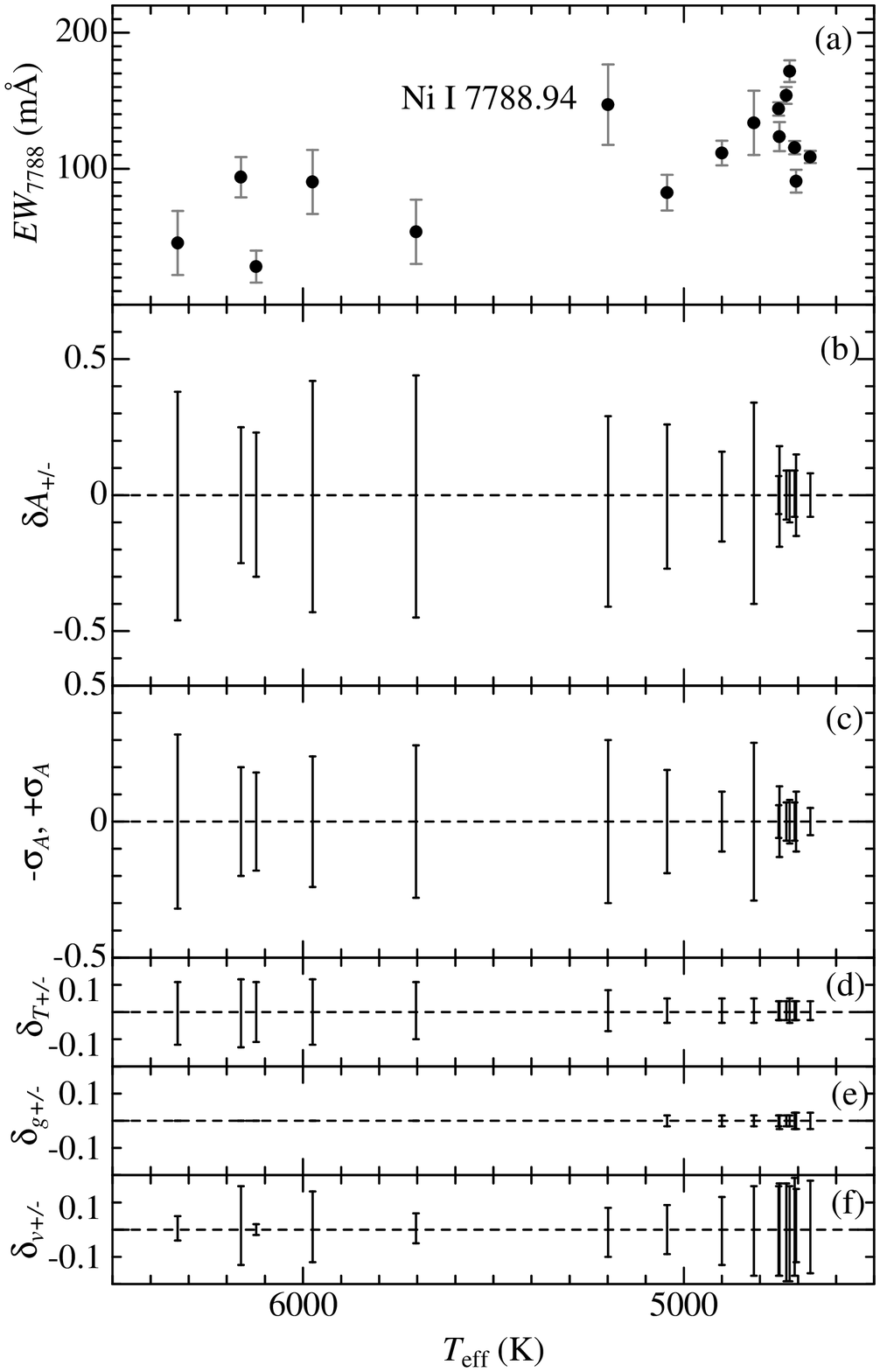}
\end{center}
\caption{
Quantities related to Ni abundance errors plotted against $T_{\rm eff}$. 
While panel (a) shows $EW^{\rm Ni}_{7788}$ (equivalent width of 
Ni~{\sc i} 7788.94), the meanings of panels (b)--(f) are the same as 
in figure 8. The signs of $\delta$'s are $\delta_{T+}>0$, $\delta_{T-}<0$,
$\delta_{g+}>0$, $\delta_{g-}<0$, $\delta_{v +}<0$, and
$\delta_{v -}>0$.
}
\end{figure}

\setcounter{figure}{10}
\begin{figure}
\begin{center}
  \FigureFile(100mm,140mm){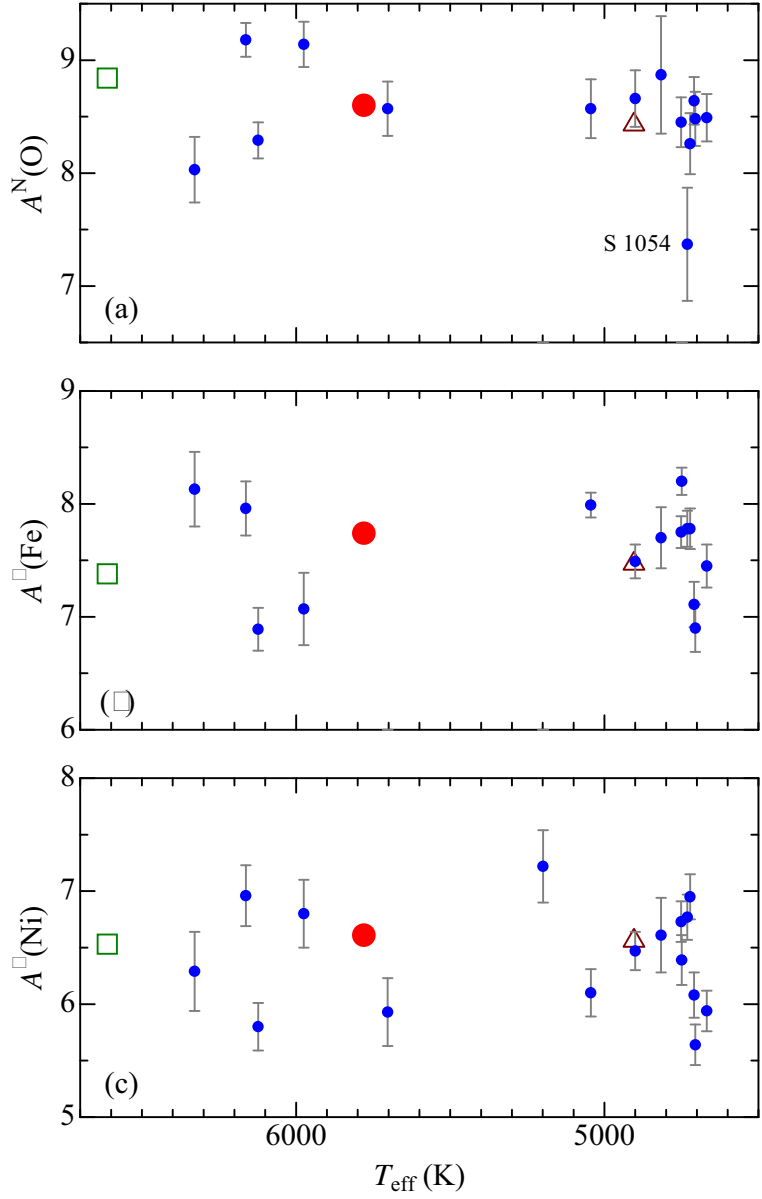}
\end{center}
\caption{
Finally obtained abundances of (a) O, (b) Fe, and (c) Ni, plotted against $T_{\rm eff}$
The results for M~67 stars are shown in small symbols (blue filled circles), where the 
attached error bars are $\pm \sqrt{\sigma_{A}^{2} + \delta_{Tgv}^{2}}$ (cf. subsection 5.4),
while those for Pollux, Procyon, and the Sun are presented in large symbols 
(brown open triangle, green open square, and red filled circle, respectively). 
}
\end{figure}

\end{document}